\documentclass[preprint,prc,letterpaper,11pt,twoside,tightenlines,nofootinbib,showpacs]{revtex4-1}
\usepackage{amsmath,amssymb,bm}
\usepackage{color}
\usepackage{graphicx}
\usepackage{epstopdf}

\newcommand{\dd}{\mathrm{d}}
\newcommand{\MeV}{\mathrm{MeV}}
\newcommand{\GeV}{\mathrm{GeV}}
\newcommand{\TeV}{\mathrm{TeV}}

\newcommand{\bvec}[1]{\ensuremath{\boldsymbol{#1}}}
\newcommand{\erw}[1]{\ensuremath { %
    \left \langle {#1} \right \rangle}}

\begin{document}

%\preprint{APS/123-QED}

\title{Elliptic flow and nuclear modification factors of $D$-mesons at FAIR in a Hybrid-Langevin approach}

\author{Thomas Lang$^{1,2}$} \author{Hendrik van Hees$^{1,2}$} \author{Jan Steinheimer$^{3}$}
\author{Marcus Bleicher$^{1,2}$}
\affiliation{
  $^{1}\,$Frankfurt Institute for Advanced Studies
  (FIAS), Ruth-Moufang-Str. 1, 60438 Frankfurt am Main, Germany}
\affiliation{
  $^{2}\,$Institut f\"ur Theoretische Physik, Johann Wolfgang
  Goethe-Universit\"at, Max-von-Laue-Str. 1, 60438 Frankfurt am Main,
  Germany}
\affiliation{
  $^{3}\,$Lawrence Berkeley National Laboratory, 1 Cyclotron Road, Berkeley, CA 94720, USA}

\date{\today}

\begin{abstract}
  The Compressed Baryonic Matter (CBM) experiment at the Facility for
  Anti-proton and Ion Research (FAIR) will provide new possibilities for
  charm-quark ($D$-meson) observables in heavy-ion collisions at low
  collision energies and high baryon densities. To predict the
  collective flow and nuclear modification factors of charm quarks in
  this environment, we apply a Langevin approach for the transport of
  charm quarks in the UrQMD (hydrodynamics + Boltzmann) hybrid
  model. Due to the inclusion of event-by-event fluctuations and a full
  (3+1) dimensional hydrodynamical evolution, the UrQMD hybrid approach
  provides a realistic evolution of the matter produced in heavy-ion
  collisions.

  As drag and diffusion coefficients we use a resonance approach for
  elastic heavy-quark scattering and assume a decoupling temperature of
  the charm quarks from the hot medium of $130\, \MeV$. Hadronization of
  the charm quarks to $D$-mesons by coalescence is included. 
  Since the initial charm-quark distribution at FAIR is unknown, 
  we utilize two different initial charm-quark distributions in our approach 
  to estimate the uncertainty of these predictions. 
  We present
  calculations of the nuclear modification factor, $R_{AA}$, as well as
  for the elliptic flow, $v_2$, in Pb+Pb collisions at $E_{lab}=25\,\text{AGeV}$. 
  The different medium modifications of $D$-mesons and
  $\bar{D}$-mesons at high baryon-chemical potential are explored by
  modified drag- and diffusion-coefficients using the corresponding 
  fugacity factor. Here we find a considerably larger medium
  modification for $\overline{\mathrm{D}}$- than for $D$-mesons.
\end{abstract}

\maketitle

%\tableofcontents

\section{Introduction}
\label{sec:Langevin}
\markright{Thomas Lang, Hendrik van Hees, Jan Steinheimer, and Marcus
  Bleicher. Medium modification of charm quarks at FAIR in a Langevin
  approach. }

The new Facility for Anti-proton and Ion Research (FAIR) at GSI in
Darmstadt, Germany, will provide novel possibilities of probing strongly
interacting matter at high net-baryon densities
\cite{Friman:2011zz}. This extreme matter is reached by colliding heavy ions 
(i.e., gold or lead) at high collision energies. So far the most
well-known experiments have been focused on increasing the collision
energy (i.e., $\sqrt {s_{NN}}=200\; \GeV$ at RHIC and $\sqrt
{s_{NN}}=2.76\; \TeV$ at LHC). While these experiments already gave and
still give great insights to the phase of deconfined matter, where
quarks and gluons are quasi free (the Quark Gluon Plasma, QGP), the
region of high net-baryon density in the QCD phase diagram has not been
explored in detail so far. Nevertheless, the high-density
region is extraordinarily interesting for strong-interaction physics, 
because the region around the critical point of the QCD phase transition to
the QGP is expected to be covered with experiments at high baryon
chemical potential.

On the theory side a multitude of (potential) signatures and properties
of the QCD (phase) transition and the QGP have been predicted
\cite{Adams:2005dq,Adcox:2004mh,Muller:2012zq}. Some of these signatures
are related to heavy quarks \cite{Matsui:1986dk}. Since heavy quarks
(i.e., c and b quarks) are produced in the primordial hard collisions of
the nuclear reaction, they probe the created medium during its entire
evolution process. When the system cools down they hadronize and the
resulting heavy-flavour mesons can be detected. Therefore, heavy-quark
observables provide new insights into the interaction processes within
the hot and dense medium. Two of the most interesting observables are
the elliptic flow, $v_2$, and the nuclear modification factor, $R_{AA}$,
of (open) heavy-flavour mesons.  Experimentally, the nuclear modification
factor shows a large suppression of the open heavy-flavour particles'
spectra at high transverse momenta ($p_T$) compared to the findings in
pp collisions at RHIC and LHC. This indicates a high degree of
thermalization of the heavy quarks with the bulk medium consisting of
light quarks and gluons and, 
% perhaps,
at the later stages of the fireball evolution, the hot and dense hadron
gas. The measured large elliptic flow, $v_2$, of open-heavy-flavour
mesons at the various heavy-ion facilities underlines that heavy quarks
take part in the collective motion of the bulk medium. A quantitative
analysis of the degree of thermalization of heavy quarks in terms of the
underlying microscopic scattering processes thus leads to an
understanding of the mechanisms underlying the large coupling strength
of the QGP and the corresponding transport properties.

In this paper we explore the medium modification of heavy-flavour $p_T$
spectra at FAIR, using a hybrid model, consisting of the
Ultra-relativistic Quantum Molecular Dynamics (UrQMD) model
\cite{Bass:1998ca,Bleicher:1999xi} and a full (3+1)-dimensional ideal
hydrodynamical model \cite{Rischke:1995ir,Rischke:1995mt} to simulate
the bulk medium (for details of the hybrid approach see
\cite{Petersen:2006vm,Steinheimer:2007iy,Li:2008qm,Petersen:2008dd,Petersen:2008dd,Petersen:2009vx,Steinheimer:2009nn,Petersen:2010cw}).
The heavy-quark propagation in the medium is described by a relativistic
Langevin approach. Similar studies at higher energies have recently been
performed in a thermal fireball model with a combined
coalescence-fragmentation approach
\cite{vanHees:2007me,vanHees:2007mf,Greco:2007sz,vanHees:2008gj,Rapp:2008fv,
  Rapp:2008qc,Rapp:2009my}, in an ideal hydrodynamics model with a
lattice-QCD EoS \cite{He:2012df,He:2012xz}, in a model from Kolb and
Heinz \cite{Aichelin:2012ww}, in the BAMPS model
\cite{Uphoff:2011ad,Uphoff:2012gb}, the MARTINI model
\cite{Young:2011ug} as well as in further studies and model comparisons
\cite{Moore:2004tg,Vitev:2007jj,Gossiaux:2010yx,Gossiaux:2011ea,Gossiaux:2012th}.

Especially at moderate beam energies pQCD based models like BAMPS or
MARTINI can not be applied. Also 2+1 dimensional hydrodynamics,
e.g. Heinz-Kolb-Hydrodynamics, is not justified due to the strong
three-dimensional expansion. Therefore the UrQMD hybrid model provides
a major step forward as compared to simplified expanding fireball models
employed so far. It provides a realistic and well established
background, including event-by-event fluctuations and has been shown to
very well describe many collective properties of relativistic heavy-ion
collisions.

This study is the first Langevin simulation for heavy quarks to be
processed at FAIR energies. Here, a special difficulty is due to the
high baryon densities as already mentioned above. To account for these
high baryon densities we implement an optional fugacity factor for our
drag and diffusion coefficients. Within the here employed
resonance-scattering model
\cite{vanHees:2004gq,vanHees:2005wb,Lang:2012cx} this fugacity factor
leads to a stronger medium modification of $\overline{D}$- compared to $D$-mesons.

\section{Description of the model}

The UrQMD hybrid model combines the advantages of transport theory and
(ideal) fluid dynamics \cite{Petersen:2008dd}. It uses fluctuating
initial conditions \cite{Petersen:2008dd}, generated by the UrQMD model
\cite{Bass:1999tu,Dumitru:1999sf}, for a full (3+1) dimensional ideal
fluid dynamical evolution, including the explicit propagation of the
baryon current. After a Cooper-Frye transition back to the transport
description, the freeze-out of the system is treated gradually within
the UrQMD approach. The hybrid model has been successfully applied to
describe particle yields and transverse dynamics from AGS to LHC
energies
\cite{Petersen:2008dd,Steinheimer:2007iy,Steinheimer:2009nn,Petersen:2010cw,Petersen:2011sb}
and provides therefore a reliable basis for the flowing bulk medium.

The equation of state employed for the present calculations includes
quark and gluonic degrees of freedom coupled to a hadronic
parity-doublet model \cite{Steinheimer:2011ea}. It has a smooth
crossover at low baryon densities between an interacting hadronic system
and a quark gluon plasma and a (second) first order transition towards
higher baryon densities.  The thermal properties of the EoS are in
agreement with lattice QCD results at vanishing baryon density. For the
present study at FAIR energy we rely on the extrapolation to high baryon
densities as described in \cite{Steinheimer:2011ea}.

The diffusion of a heavy quark in a medium consisting of light quarks
and gluons can be described with help of a Fokker-Planck equation
\cite{Svet88,MS97,Moore:2004tg,vanHees:2004gq,HGR05a,vanHees:2007me,Gossiaux:2008jv,He:2011yi,Lang:2012cx,Lang:2012yf,Lang:2012vv}
as an approximation of the collision term of the corresponding Boltzmann
equation. It can be mapped into an equivalent stochastic Langevin
equation, suitable for numerical simulations. In the relativistic realm
such a Langevin process reads
\begin{equation}
\begin{split}
\label{lang.1}
\dd x_j &= \frac{p_j}{E} \dd t, \\
\dd p_j &= -\Gamma p_j \dd t + \sqrt{\dd t} C_{jk} \rho_k.
\end{split}
\end{equation}

Here $\dd t$ is the time step in the Langevin calculation, $\dd x_j$ and
$\dd p_j$ are the coordinate and momentum changes in each time-step, $m$
is the heavy-quark mass, $E=\sqrt{m^2+\bvec{p}^2}$, and $\Gamma$ is the
drag or friction coefficient.  The covariance matrix, $C_{jk}$, of the
fluctuating force is related with the diffusion coefficients. Both
coefficients depend on $(t,\bvec{x},\bvec{p})$ and are defined in the
(local) rest frame of the fluid. The $\rho_k$ are Gaussian-normal
distributed random variables, i.e., its distribution function reads
\begin{equation}
\label{lang.2}
P(\bvec{\rho}) = \left (\frac{1}{2 \pi} \right)^{3/2} \exp
\left(-\frac{\bvec{\rho}^2}{2} \right ).
\end{equation}
The fluctuating force obeys
\begin{equation}
\label{lang.3}
\erw{F_j^{(\text{fl})}(t)}=0, \quad \erw{F_j^{(\text{fl})}(t)
  F_k^{(\text{fl})}(t')} = C_{jl} C_{kl} \delta(t-t').
\end{equation}
We use \cite{Dunkel-Haenggi:2008} 
\begin{equation}
\label{lang.4}
C_{jk} =C_{jk}(t,\bvec{x},\bvec{p}+\dd \bvec{p}).
\end{equation}

The drag and diffusion coefficients for the heavy quark propagation
within this framework are taken from a resonance approach
\cite{vanHees:2004gq,vanHees:2005wb}. It is a non-perturbative model,
where the existence of $D$-meson like resonances in the QGP phase is
assumed.

The initial production of charm quarks in our approach is based on a
space-time resolved Glauber approach. For the realization of the initial
collisions for the charm quark production points we use the UrQMD
model. First, an UrQMD run with straight trajectories is performed. Here, 
only elastic $0^\circ$ scatterings between the colliding nuclei are
carried out and the nucleon-nucleon collision space-time coordinates are
stored (see \cite{Spieles:1999kp}).  These coordinates are used in a
second, full UrQMD run as probability distribution for the production space-time coordinates for the
charm quarks.

The initial momentum distribution of the produced charm quarks at FAIR
energy is not well known, due to the lack of experimental measurements
of $D$-meson production at these low energies. 
To account for this uncertainty, we utilize two different momentum distributions for the initial state of $D$ and $\bar{D}$, 
a parametrization used in HSD calculations and the distribution generated by the PYTHIA model. 

The HSD parametrization (derived from higher energies) is taken from a fit to the $D$-meson distribution 
in pp collisions at $25\,\text{AGeV}$ \cite{Cassing:2000vx}. 
As fitting function we use
\begin{equation}
\label{HSD}
\frac{\text{d}N}{\text{d}p_T} =\frac{C_1}{\left(1+A_1\cdot p_T^2\right)^{A_2}},
\end{equation}
with the coefficients $A_1=\,0.870/\text{GeV}^2$ and $A_2=\,3.062$. $C_1$ is an 
arbitrary normalization constant with the unit $1/\text{GeV}$. 

The PYTHIA parametrization is extracted from a fit to pp collisions generated by  
PYTHIA \cite{Sjostrand:2006za} at $E_{lab}=25\,\text{AGeV}$. Here we use as fitting function 
\begin{equation}
\label{PYTHIA}
\frac{\text{d}N}{\text{d}p_T} = C_2\,p_T^{B_1}\,\text{exp}\left(-\frac{(p_T-B_2)^2}{2B_3^2}\right),
\end{equation}
with the coefficients $B_1=\,1.144$, $B_2=\,-0.586\,\text{GeV}$ and $B_3=\,0.646\,\text{GeV}$. 
$C_2$ is an arbitrary normalization constant with the unit $\text{GeV}^{-B_1-1}$. 
In the following we normalize the distributions to $\int \text{d}p_T\frac{\text{d}N}{\text{d}p_T}=1$. 
These two parametrizations are utilized as our initial charm-quark distributions. 
Due to the low collision energy, charm quark production in pp collisions is kinematically only allowed up to a transverse momentum of  
$p_T\approx 2\,\text{GeV}$. Although the production cut-off for charm quarks in Pb+Pb collisions 
might be somewhat higher due to nuclear effects, 
our calculations above $p_T\approx 1.5\,\text{GeV}$ should be interpreted with caution only. 
This is especially true for our $R_{AA}$ calculations. 

Starting with these charm-quark distributions as initial condition we 
propagate the charm quarks on straight lines until 
the hydro start condition is fulfilled. For the start condition we use 
$t_{\text{start}} = 2R/\sqrt{\gamma_{\text{CM}}^2 -1}$,
i.e., after the two Lorentz-contracted nuclei have passed through each
other ($\gamma_{\text{CM}}$ is the centre-of-mass-frame Lorentz factor,
and $R$ is the radius of the nucleus). For the Langevin calculation we
use the UrQMD/hydro's cell velocities, cell temperature and the size of
the time-step for the calculation of the momentum transfer, propagating
all quarks independently.  The charm-quark propagation is terminated by
hadronization into $D$-mesons, via quark-coalescence
\cite{Lang:2012cx,Lang:2012yf,Lang:2012vv}.

\section{Results}

\begin{figure}[h]
\begin{minipage}[b]{0.45\textwidth}
\includegraphics[width=1\textwidth]{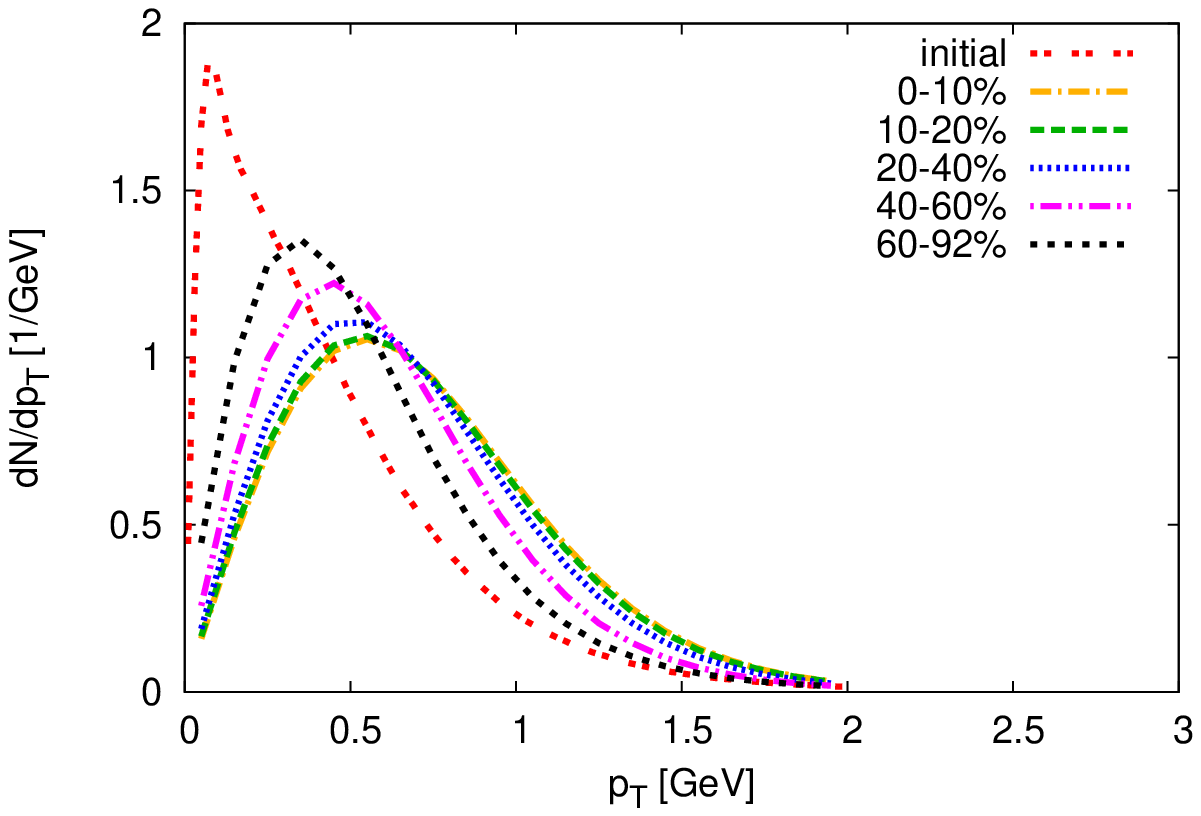}
\end{minipage}
\hspace{5mm}
\begin{minipage}[b]{0.45\textwidth}
\includegraphics[width=1\textwidth]{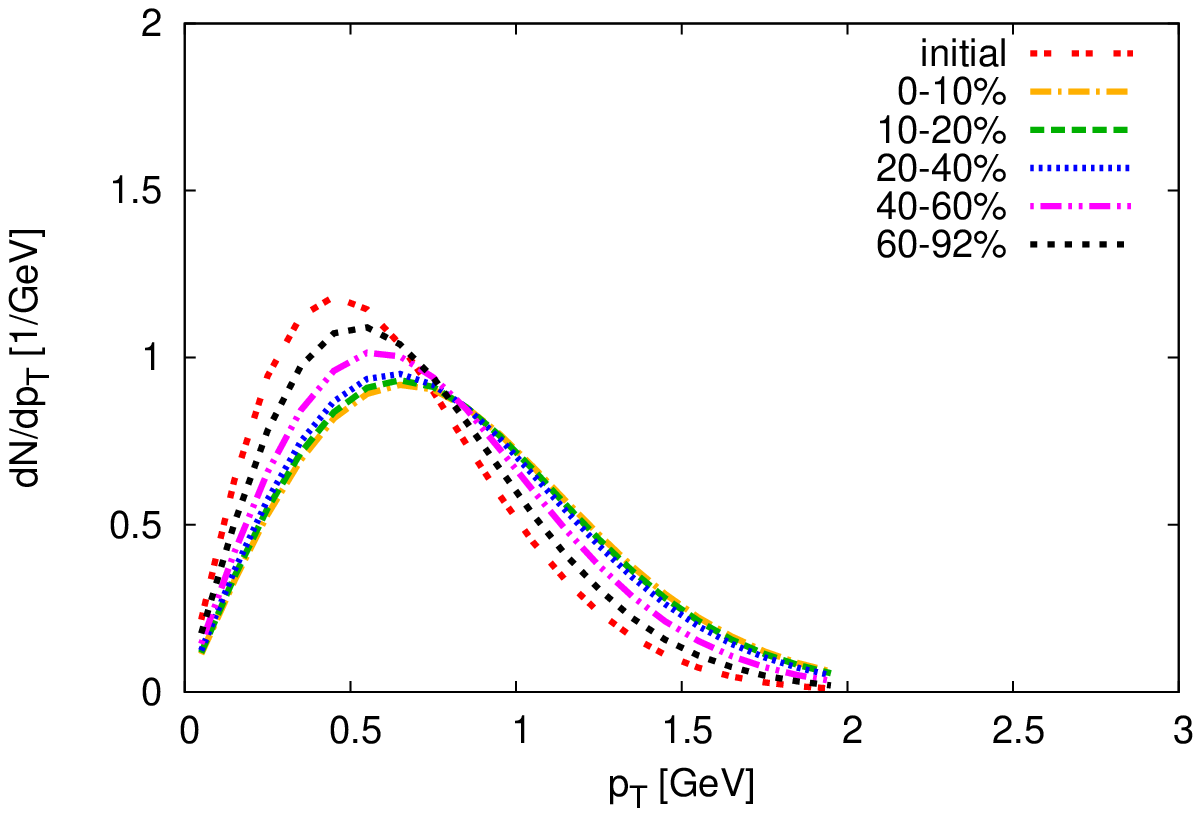}
\end{minipage}
\caption{(Color online) The initial and final normalized $p_T$ spectra of $D$-mesons in
  Pb+Pb collisions at $25\,\text{AGeV}$ for different centrality bins. 
  The left plot shows our calculation for the HSD initial-state parametrization, the right plot 
  shows it for the PYTHIA initial-state parametrization. 
  We use a rapidity cut of $|y|<0.35$.}
\label{ptspectra}
\end{figure}

Let us start with the initial and final $D$-meson $p_T$-spectra at five different centrality 
bins for Pb+Pb collisions at $E_{\text{lab}}=25\,\text{AGeV}$, as shown in Fig.\ \ref{ptspectra}. 

The initial $D$-meson distributions are for both initial state assumptions much softer as
compared to $D$-meson $p_T$-spectra observed in pp collisions at RHIC or LHC energies.  
In the HSD parametrization most charm quarks are at very low transverse momenta,  
and the initial distribution falls down according to a power law. 
The PYTHIA parametrization has a maximum at $p_T\approx 0.5\,\text{GeV}$. 
At high $p_T$ however, it decreases faster than the HSD parametrization due to the exponential in 
the distribution function (equation \ref{PYTHIA}). 

The final distributions show a thermalization of the charm quarks. 
The propagation of the charm quarks in the hot medium and the coalescence
mechanism drag low-$p_T$ particles to higher $p_T$ bins. This drag is
more pronounced at high collision centralities.

\begin{figure}[h]
\begin{minipage}[b]{0.45\textwidth}
\includegraphics[width=1\textwidth]{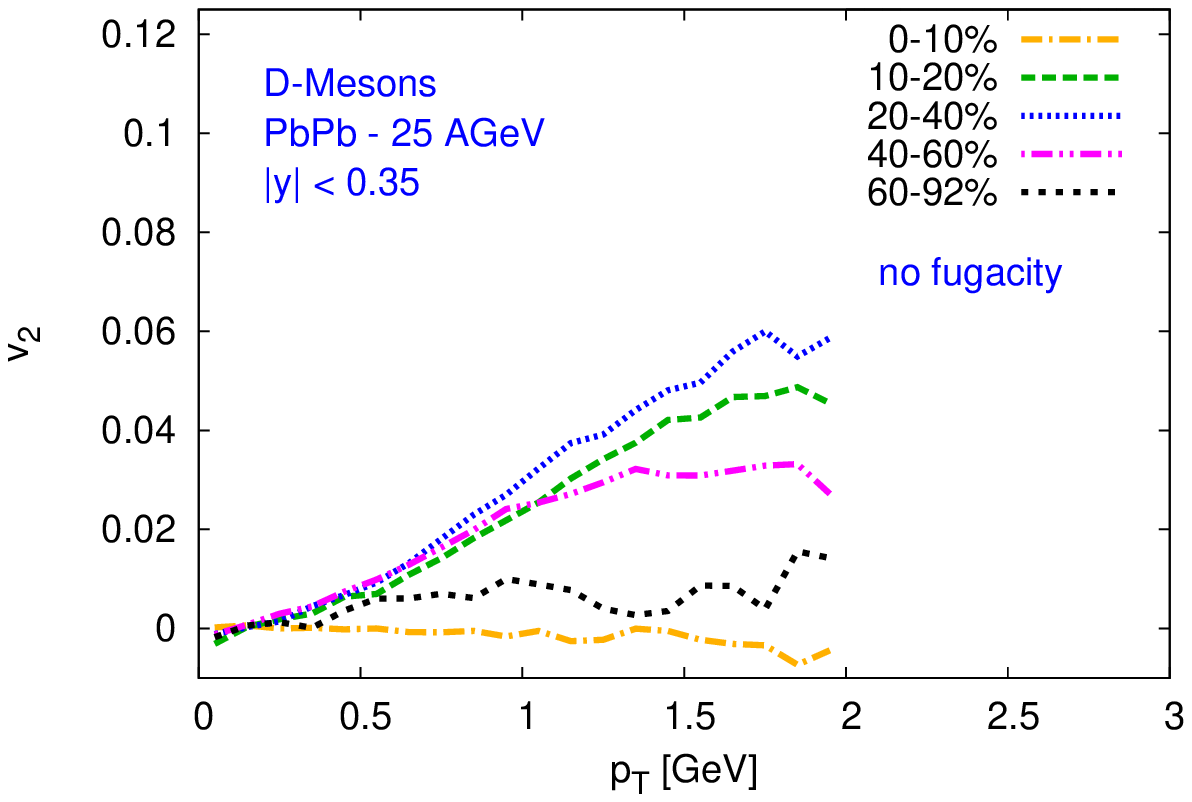}
\end{minipage}
\hspace{5mm}
\begin{minipage}[b]{0.45\textwidth}
\includegraphics[width=1\textwidth]{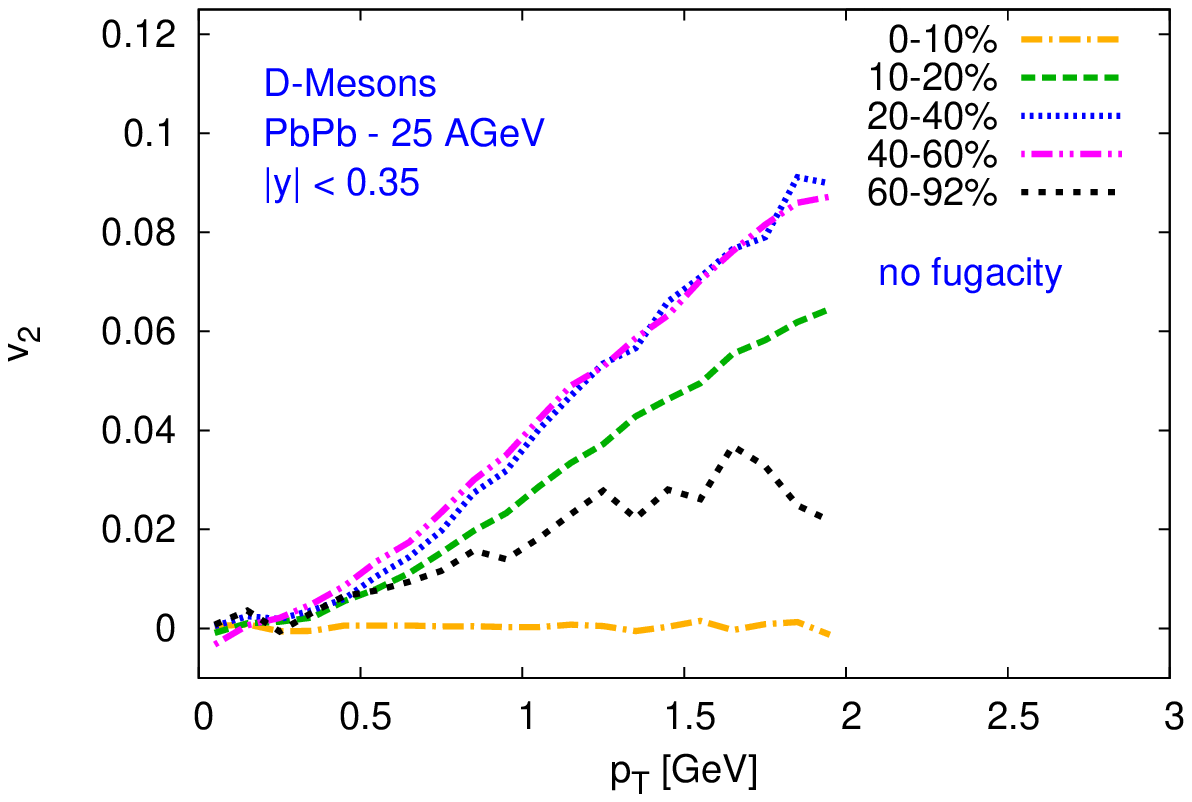}
\end{minipage}
\caption{(Color online) Elliptic flow, $v_2$, of $D$-mesons in
  Pb+Pb collisions at $25\,\text{AGeV}$ for different centrality bins. 
The left plot shows our calculation for the HSD initial-state parametrization, the right plot 
  shows it for the PYTHIA initial-state parametrization. 
  We use a rapidity cut of $|y|<0.35$.}
\label{FlowFAIR4}
\end{figure}

\begin{figure}[h]
\begin{minipage}[b]{0.45\textwidth}
\includegraphics[width=1\textwidth]{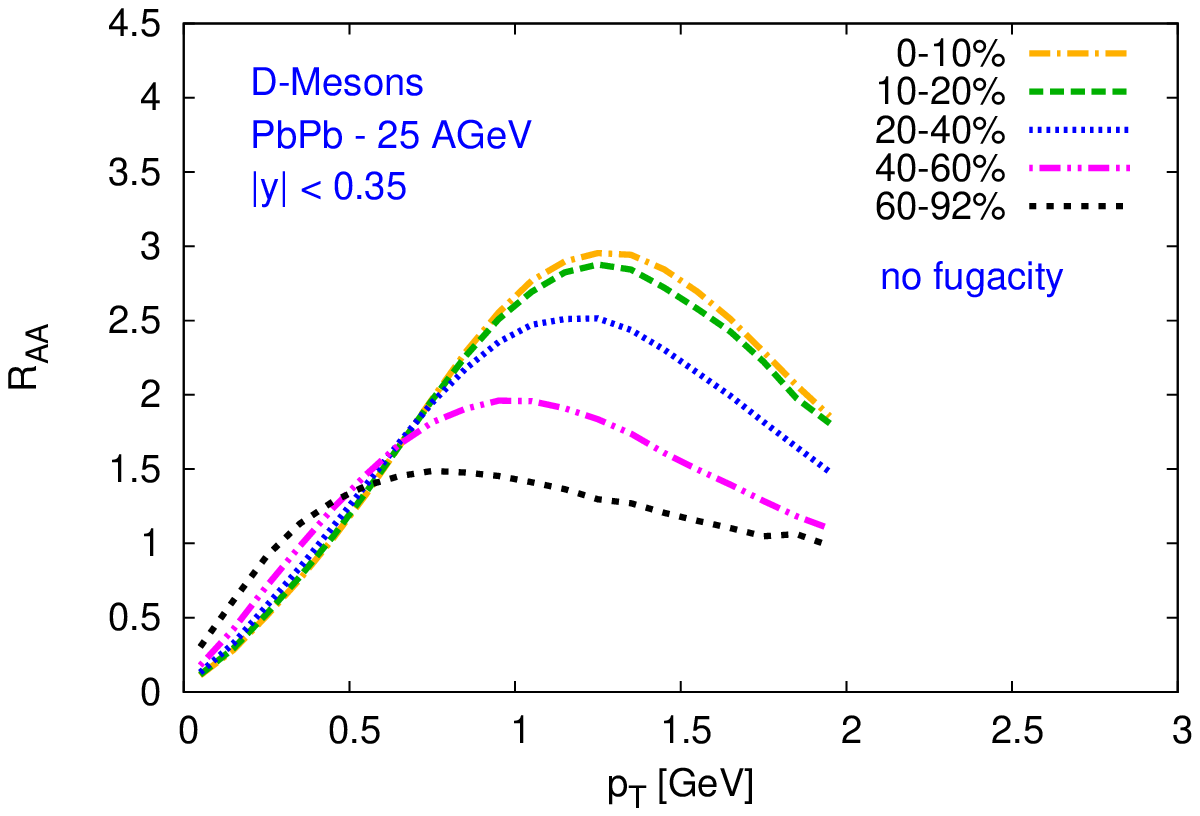}
\end{minipage}
\hspace{5mm}
\begin{minipage}[b]{0.45\textwidth}
\includegraphics[width=1\textwidth]{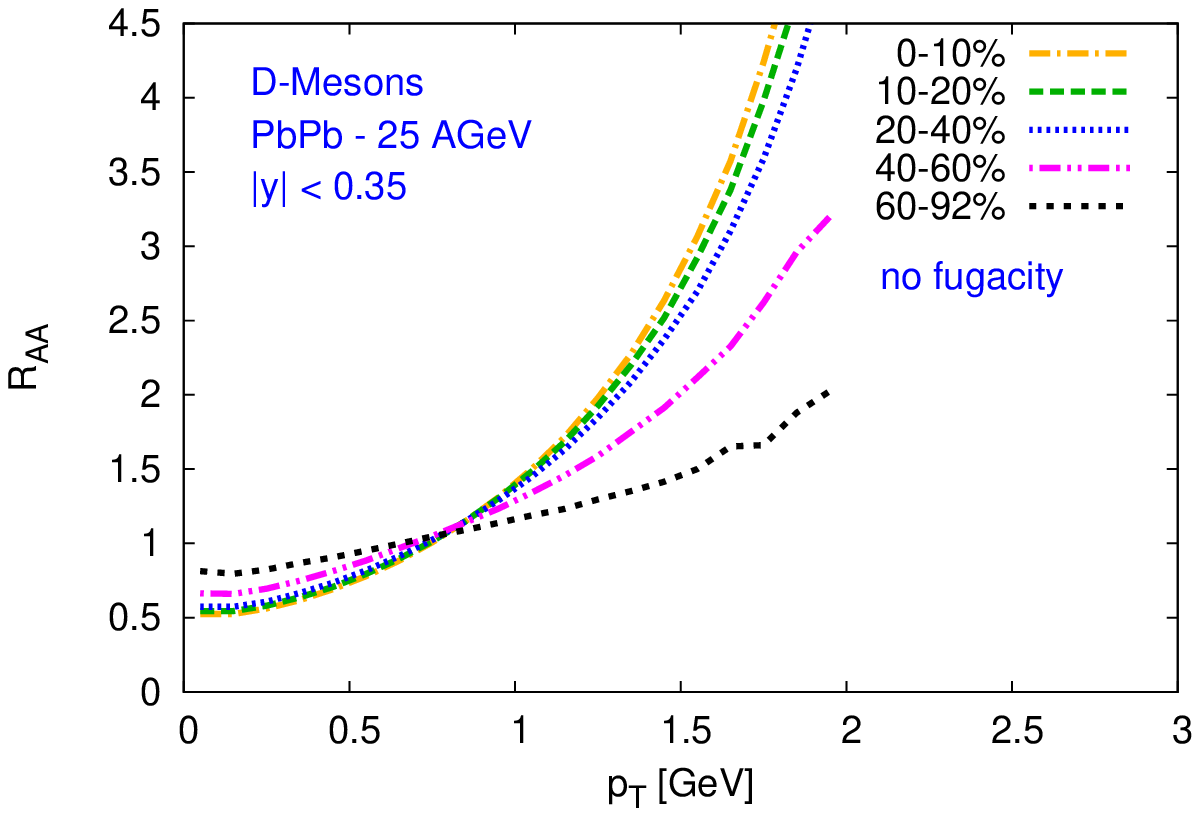}
\end{minipage}
\caption{(Color online) Nuclear modification factor, $R_{AA}$, of $D$-mesons in
  Pb+Pb collisions at $25\,\text{AGeV}$ for different centrality bins. 
The left plot shows our calculation for the HSD initial-state parametrization, the right plot 
  shows it for the PYTHIA initial-state parametrization. 
  We use a rapidity cut of $|y|<0.35$. }
\label{RAAFAIR4}
\end{figure}

In the following we show the elliptic flow, $v_2$, and the nuclear
modification factor, $R_{AA}$, of $D$-mesons (for a comparison with the light quark hadron $v_2$, the reader is referred to \cite{Petersen:2006vm}).  Our results are depicted in
Fig.\ \ref{FlowFAIR4}, for the elliptic flow, and in Fig.\ \ref{RAAFAIR4},
for the nuclear modification factor.

The strongest elliptic flow in our calculation can be observed for a
medium centrality range of $\sigma
/\sigma_{\mathrm{tot}}=20\%$-$40\%$. It reaches up to around $6\%$ in case of the HSD initial-state parametrization, 
which is approximately half of the elliptic flow which develops at RHIC energies. 
In case of the PYTHIA parametrization the flow is considerably larger and reaches up to $10\%$. 
In both cases the flow for very central and peripheral collisions is, as
expected, small. The higher maximal $v_2$ values with PYTHIA initial conditions 
are due to the fact that high-$p_T$ charm quarks need more interactions with the medium 
to reach a high $p_T$ due to the softer $p_T$ spectrum as compared to the HSD intial state. 

Let us now turn to the nuclear modification factor, $R_{AA}$, for 
the case of the HSD parametrization (Fig.\ \ref{RAAFAIR4}, left). 
The strongest modification is reached for central collisions and the
lowest for peripheral collisions. We also observe that the highest
$R_{AA}$ values move to higher $p_T$ for more central collisions.
 
Compared to the nuclear modification factor at RHIC and LHC energies
\cite{Lang:2012cx,Lang:2012yf,Lang:2012vv} the medium modification is extremely
large and qualitatively different. On the one hand this effect is due to
the initial charm-quark distribution (HSD initial-state parametrization) that we used as one assumption. 
It is much softer than at RHIC or LHC energies. Therefore it drops off very fast towards
higher $p_T$. Thus the drag of low $p_T$ particles to higher $p_T$ has a
large relative influence on the $R_{AA}$ at higher $p_T$ ranges.  On the
other hand the medium modification of low-$p_T$ heavy quarks seems to be
stronger at FAIR energies. As one can see in Fig.\ \ref{ptspectra} (left) a
big fraction of the $D$-mesons at low $p_T$ is shifted to higher
$p_T$.  This effect can be explained by the low momenta, also in
longitudinal direction, of the heavy quarks and the slow medium
evolution at FAIR energies. Therefore the charm quarks stay a long time
in the medium and can ``heat up'' due to the diffusion in our Langevin
calculation. 

The nuclear modification factor using the PYTHIA initial-state parametrization (Fig.\ \ref{RAAFAIR4}, right) 
looks completely different. At low $p_T$ the PYTHIA parametrization has a similar shape 
as the medium modified $\text{d}N/\text{d}p_T$ distribution for the HSD calculation (Fig.\ \ref{ptspectra}). 
Therefore the medium modification at low $p_T$ is not as strong as for the former case. 
For higher $p_T$ the $R_{AA}$ rises drastically, especially for central collisions. 
The reason is the initial parametrization in equation \ref{PYTHIA}. 
At high $p_T$-values it drops faster than a thermal distribution. 
Therefore, the thermalization of charm quarks drags some of the charm quarks 
to high-$p_T$ bins that are strongly suppressed (or even forbidden) 
by energy conservation in the initial state.  \\

\begin{figure}[h]
\includegraphics[width=0.7\textwidth]{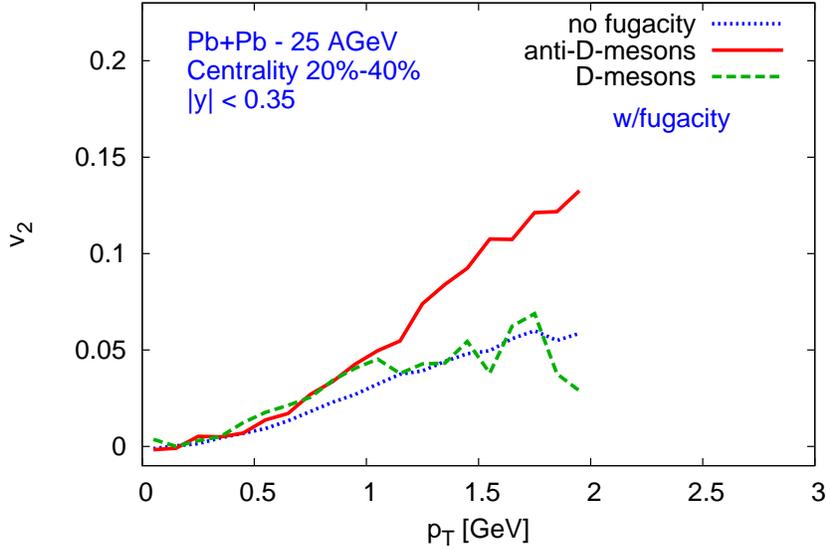}
\caption{(Color online) Elliptic flow, $v_2$, of $D$-mesons and
  $\bar{D}$-mesons in Pb+Pb collisions at $25\,\text{AGeV}$ using the HSD initial-state parametrization and applying 
  fugacity factors.
  We use a rapidity cut of $|y|<0.35$. The results above $p_T\approx 2\,\text{GeV}$ should be interpreted with caution only.}
\label{FlowFAIRFug}
\end{figure}

\begin{figure}[h]
\includegraphics[width=0.7\textwidth]{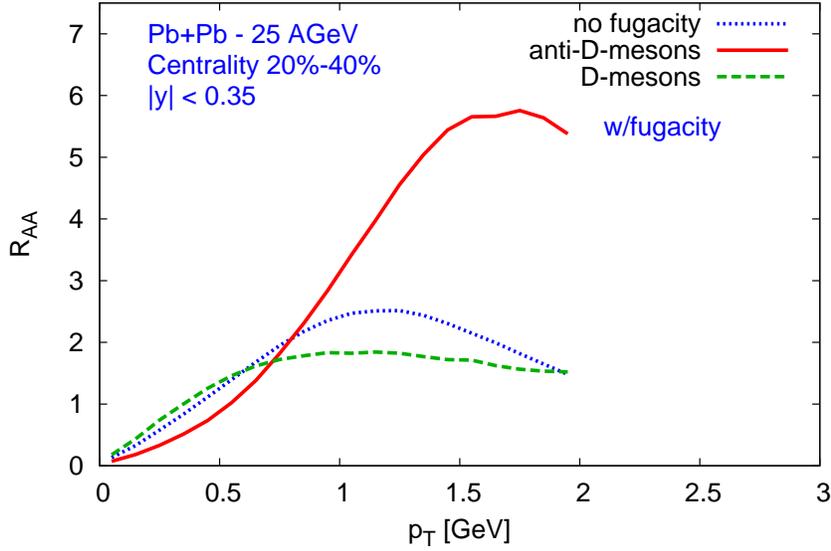}
\caption{(Color online) Nuclear modification factor, $R_{AA}$, of $D$-mesons and
  $\bar{D}$-mesons in Pb+Pb collisions at $25\,\text{AGeV}$ using the HSD initial-state parametrization and applying 
  fugacity factors.  We use a rapidity cut of $|y|<0.35$. The results above $p_T\approx 2\,\text{GeV}$ should be interpreted with caution only.}
\label{RAAFAIRFug}
\end{figure}

In the next step we include fugacity factors in our calculation to
account for the high baryon chemical potential at FAIR-energies. 
Therefore we multiply the
anti-charm drag- and diffusion-coefficients by $e^{\mu_B /T}$ and the
charm coefficients by $e^{-\mu_B /T}$.  Here $\mu_B$ is the baryon
chemical potential of the surrounding quarks and $T$ is the local
temperature of the medium. 
As initial charm-quark distribution we used the HSD parametrization of equation (\ref{HSD}). 
Fig.\ \ref{FlowFAIRFug} shows our results for
the elliptic flow and Fig.\ \ref{RAAFAIRFug} for the nuclear modification
factor.

As one can see the inclusion of fugacity factors changes the results
substantially.  The elliptic flow for $\overline{\mathrm{D}}$-mesons reaches up to
$15\%$ and also the nuclear modification factor changes strongly.  The
difference between $D$-mesons and $\bar{D}$-mesons is clearly visible.  If we
have a look at the difference between the $D$-mesons and the calculation
neglecting fugacity-factors we realize that the difference is much
smaller than for $\overline{\mathrm{D}}$-mesons.  This small difference
is not due to a small difference of the coefficients used, but to the
role of the coalescence mechanism that accounts for the overwhelming
fraction of the elliptic flow of $D$-mesons if the coefficients are small.

Also in case of the nuclear modification factor, $R_{AA}$, one observes a
strong difference between $D$-mesons and $\overline{\mathrm{D}}$-mesons.
Overall the medium modification is considerably stronger than at RHIC
and LHC energies. We relate this to two different effects. The first is
due to the very soft initial momentum distribution of the charm quarks.
Therefore a small change of the $R_{AA}$ at low $p_T$ can result in a 
substantial $R_{AA}$ change at higher $p_T$. The second effect comes from the
slower bulk-medium evolution at FAIR energies compared to RHIC and LHC
energies. This slower evolution results in a longer time that the charm
quarks stay in the medium and thus in a stronger drag/diffusion towards
thermalization. In the $R_{AA}$ this results in a strong suppression at
low $p_T$ due to the resulting ``heat-up'' of the charm quarks.

We should mention that the difference seen between $D$-mesons and
$\overline{\mathrm{D}}$-mesons is sensitive to the model used to
calculate the drag- and diffusion-coefficients. In case of the
$T$-Matrix approach \cite{vanHees:2007me} this difference should not
arise. Therefore $D$-meson measurements at FAIR can provide an excellent
test for a confirmation or rejection of different heavy-quark-coupling
mechanisms to the QGP.

\section{Summary}

In this letter we have explored the medium modification of $D$-meson spectra at FAIR energies. 
While the elliptic flow 
is on the same order as compared to higher energies, the $R_{AA}$ shows 
a strong modification. This modification depends strongly on the initial-state 
parametrization of the charm-quark momentum distribution. For $R_{AA}$ we observe for the HSD initial-state 
parametrization (neglecting the energy cut-off for $D\bar{D}$ production) 
a similar shape as for higher energies, while for the 
PYTHIA initial-state parametrization $R_{AA}$ rises monotonously with $p_T$ due to the sharp drop-off of the 
transverse momentum spectra in pp. 

We also included fugacity factors in the calculation 
and found a substantial difference in the medium modification of $D$-mesons and $\bar{D}$-mesons. 
However, this difference only appears in case of utilizing the resonance model coefficients and 
therefore provides an excellent possibility to disentangle models for the calculation of drag and diffusion 
coefficients.

\section*{ACKNOWLEDGMENTS}

We are grateful to the Center for Scientific Computing (CSC) and the
LOEWE-CSC at Frankfurt for providing computing resources. T.~Lang
gratefully acknowledges support from the Helmholtz Research School on
Quark Matter Studies. This work is supported by the Hessian LOEWE
initiative through the Helmholtz International Center for FAIR (HIC for
FAIR). J.~S. acknowledges a Feodor Lynen fellowship of the Alexander von
Humboldt foundation.  This work is supported by the Office of Nuclear
Physics in the US Department of Energy's Office of Science under
Contract No. DE-AC02-05CH11231, the GSI Helmholtzzentrum and the
Bundesministerium f{\"ur} Bildung und Forschung (BMBF) grant
No. 06FY7083.

\bibliography{bibliography}

\end{document}